\documentclass[12pt]{article}

\usepackage{graphicx}          
\usepackage{dcolumn}           
\usepackage{bm}                
\usepackage{float}             
\usepackage{amssymb}   
\usepackage{hyperref}    
\usepackage{epigraph}     

 \title{\bf Sandage versus Hubble on the reality of the expanding universe}

\author{Domingos Soares \\ \\ {Physics Department}\\
{Federal University of Minas Gerais} \\   
{Belo Horizonte, MG, Brasil} }

\date{May 31, 2006}

\begin{document}
\maketitle

\hfill {\it `We are certainly not to relinquish the evidence of experiments} 

\hfill {\it for the sake of dreams and vain fictions of our own devising.' }\\

\hfill Mathematical Principles of Natural Philosophy,

\hfill  Book III --- I. Newton, 1687\\

\begin{abstract}
A critical reading of Lubin \& Sandage's 2001 paper on the Tolman
effect for the reality of the expansion of the universe clearly
reveals that Sandage is far from winning the dispute with Hubble on
the issue. After all the years, Hubble's doubt about the reality of
the expansion remains as valid as Sandage's certainty expressed in a
series of papers in the last decade.
\end{abstract}

\bigskip\bigskip\bigskip

%

\section{Introduction}

To begin with let us state clearly what are Sandage's and Hubble's
opinions on the reality of the expanding universe.

Since his discovery of the redshift-distance linear relation, Hubble
did not accept the direct interpretation of a Doppler effect as being
responsible for the spectral shifts. He was still reluctant in
accepting the reality of the expansion as late as 1953, the year of
his death (Lubin \& Sandage 2001, hereafter LS01).

Sandage, on the contrary, mainly based on his and collaborators' long
time work on the Tolman effect (in fact, since 1991, see references in
LS01), believes that the expansion of the universe is a reality.

Now, LS01's conclusion is rather {\it inconclusive}, if one sticks to basic
concepts of epistemology. After their analysis of the surface
brightness (SB) of 34 early-type galaxies is completed, they state, at
the end of \S 4.2: ``Therefore, we assert that we have either (1)
detected the evolutionary brightening directly from the $\langle SB \rangle$ 
observations {\it on the assumption that the Tolman effect exists} or (2)
confirmed that the Tolman test for the reality of the expansion is
positive, {\it provided that the theoretical luminosity correction for
evolution is real} (emphases added)."

What do they assert anyway? We shall keep for the purposes of the
present paper what they write in the abstract: ``We conclude that the
Tolman surface brightness test is consistent with the expansion to
within the {\it combined} errors of the observed $\langle SB \rangle$
 depression and the
{\it theoretical} corrections for luminosity evolution (emphases added)."
The effect may be consistent but given the conditional statements it
may not exist at all.

On the other side, Hubble's position was much more coherent, from the
scientific point of view. Although referred to as ``a reductionist
bench scientist" (LS01, \S 1.3), Hubble solely relied (mistakenly,
according to Sandage) on the interpretation of his observational data
and their accuracy. As far as we know, such a procedure --- as regular
scientific behavior --- was inaugurated by the brilliant Danish
astronomer Tycho Brahe, in the XVI century, and has proved wise and
successful beyond any doubt. But Sandage adds that besides that
{\it mistake}, Hubble used also a {\it mistaken} theory of how redshifts should
vary with distance. Why, one should ask: how could Hubble use the
correct theory if he was, to begin with, looking for the correct
theory?

The approach adopted by Sandage in his investigation of the Tolman
effect is in fact a masterpiece of {\it tautology and hermeneutical
circularity}, in spite of his clear intention of hiding it (some hints
in \S 2).

In the XXI century, Sandage still plays with $q_\circ$, $H_\circ=50$ and Mattig's
equations. When he is warned that his cosmology mates are talking now
about a Lambda-dominated universe, he reduces (a reductionist?) all of
the entire new-cosmology standard model to a simple and empty 
$q_\circ=0$ universe (quoted as {\it ``almost identical"}, see LS01, end of \S 5).

Cosmology is still a heavy-speculated field in spite of the enormous
efforts on presumable cosmology-sensitive observations. In such an
environment, scientists are not expected to make incisive statements
unless they are supported by definitely secure evidence, both on the
theoretical and experimental or observational sides. The paper under
criticism is an example of {\it the uncertain chain that links speculation
to speculation in order to confirm speculation}. The scientific
procedure is there but the scientific soul is not. In other words,
pretty and nice formal science leading to no real scientific
conclusion. That is the way LS01 should be read.

\section{The Tolman effect }
The Tolman (1930) test for the reality of the expansion, in
Friedmann-Robertson-Walker universes, predicts a (1+z)$^4$ dependence of
the surface brightness with redshift. It is formulated as
follows. Consider a source of luminosity $L_e$ at emission, located at
comoving distance $D$, on the time of reception. An observer receives
the luminosity $L_e/(1+z)^2$, dimmed by both the redshifted photons and by
time dilation on reception. The flux detected by the observer is then
given by $F = L_e/[(1+z)^24\pi D^2]$.

The observed angular size of the source, with linear size $R_e$ at
emission, is $\theta = R_e(1+z)/D$. The average surface brightness is
calculated from $\langle SB \rangle = F/(\pi\theta^2) = L_e/[4\pi^2R_e^2(1+z)^4] =
\langle SB_e \rangle/(1+z)^4$. This can be expressed in magnitudes as 
$\langle SBM \rangle = \langle SBM_e \rangle +
2.5\log(1+z)^4$, which is the usual presentation of the Tolman surface
brightness test for the reality of the expanding universe.

\section{Sandage and collaborators' inconsistencies  }
There are a number of inconsistencies in Sandage and co-workers'
approach to the Tolman test. Of course, these are often overlooked by
a biased Reader. In their last paper, LS01, the following list shows
the main drawbacks in their study.

   1) The analysis is made upon a toy model of the universe. A
   Friedmann model characterized by the deceleration parameter 
$q_\circ$, a
   Hubble constant of 50, and the classical Mattig's equations for the
   dependence of the quantities of interest on the redshift z.

   2) Three decisive proofs, presented in LS01, that the expansion is
   real are everything but {\it decisive} (see \S 1.4 and references
   therein). Two of then, the time dilation test in the light curves
   of supernovae, and the running of the blackbody radiation as a
   function of redshift are jeopardized by evolutionary effects, still
   unsolved. To accept these tests as real tests is left to anybody's
   wish. The third, namely, the so-called ``vertical normalization" of
   the background Planckian curve is justified by a conversation
   between Sandage and P.J.E. Peebles, as stated in the
   acknowledgments. Now, science needs more than {\it authoritative} 
   discussions as scientific demonstrations. Incidentally, the third
   proof is considered by LS01 (\S 1.4.3) as the definitive proof of
   the Tolman effect. One might with reason then ask: why go on
   further with the investigation?

   Speaking of authority, it is worthwhile mentioning two
   authoritative opinions on the significance of the microwave
   background radiation in cosmology. Fred Hoyle (2001) states that
\begin{quote}
    {\it  ``There is no explanation at all of the microwave background in
      the Big Bang theory. All you can say for the theory is that it
      permits you to put it in if you want to put it in. So, you look
      and it is there, so you put it in directly. It isn't an
      explanation." }
\end{quote}
   And Jean-Claude Pecker (2001) reaffirms:
\begin{quote}
    {\it  ``Actually, the 3 degree radiation, to me, has not a cosmological
      value. It is observed in any cosmology: in any cosmology you can
      predict the 3 degree radiation. So it is a proof of no cosmology
      at all, if it can be predicted of all cosmology."}
\end{quote}
   3) Section 5 of LS01 is dedicated to the tired-light {\it speculation},
   as they put it. To be fair, the discussion presented in this
   section is useless, from the scientific point of view, since it
   compares a speculation with a toy model (the Friedmann cosmology).
   Besides that, ``tired light" is in fact the name of a general
   paradigm: it is still {\it a paradigm in search of a theory} (note that
   the same epithet has been already addressed to another speculation,
   namely, Guth's inflation). Being such, there are many possible
   theories of the tired-light mechanism. It is not clear what theory
   LS01 considers, which is another weak point of their comparison. By
   the way, their intention is to compare the tired-light model with
   {\it observations}. As shown above, epistemology again teaches us that
   their approach is not valid.

   4) LS01 naturally recognizes that luminosity evolution affects both
   the observed surface brightness and the absolute magnitude of
   galaxies. But they make the crucial assumption that it does not
   affect galaxy radius (\S 3.1). Now, such an assumption is probably
   not true since the radius is calculated from the Petrossian metric
   radius, defined as the difference in magnitude between the mean
   surface brightness averaged over the area interior to a particular
   radius and the surface brightness at that radius (see \S 1.5).

   5) The calculation of the theoretical luminosity evolution from
   stellar population synthesis is also plagued with LS01's naive
   assumptions. The age as a function of redshift, T(z) (eqs. 8 and
   9), is taken from their preferred {\it toy model}. Of course, Sandage's
   stickiness to $H_\circ=50$ is somewhat alleviated here. In his (their)
   words (\S 4.1): ``For these calculations, we must use the {\it real} value
   of $H_\circ$ (emphasis added)."  One should not be surprised to know that
   his real value of $H_\circ$ is 58 km/s Mpc$^{-1}$.

   6) In \S 4.2, with the evolutionary calculation, they assume overall
   solar abundances because the metallicities of cluster galaxies are
   not strongly constrained from the observations. It is well known
   that different input metallicities onto evolutionary codes lead to
   substantial different synthesis results.

   7) In section 7, they explicitly admit two systematic uncertainties
   in the study. First, a minor technical problem in the galaxy radius
   calculation --- already contaminated by a major problem, as shown
   above --, and, second, they acknowledge the selection bias present
   in the galaxy sample. Anyway, as expected, they assure that
   ``neither of them are severe enough to jeopardize the results." We
   may otherwise simply disagree with that.

\section{Concluding remarks}
As a matter of science, the Tolman surface brightness test for the
reality of the expansion of the universe remains inconclusive.

\subsection{The contemporaneity of the doubt}
Hubble versus Sandage: two antagonized scientific attitudes. Both
scientists are confronted with the unknown and their reactions are
completely opposite to each other.  Why would Sandage's attitude be on
the wrong track? Simply because Friedmann models were at Hubble's time
as valid as arguing for an still {\it unknown} behavior of Nature as the
cause leading to the redshift-distance relation. As time went by, such
an attitude revealed itself to be more and more trustful. Nowadays,
one see that modern cosmological models --- in fact, modified Friedmann
models --- are totally unsatisfactory. One of the main desired outcomes
of modern cosmology, namely, the matter-energy content of the universe
does not conform to the real world: out of the total matter-energy
budget only 0.5\% is proved to exist from direct observations (see
summary in Soares 2002).

One might well ask: how can Sandage and collaborators make so many
weak assumptions, in the dangerous terrain of the gravely {\it unknown}, yet
be tolerated by their science mates, and at the end conclude that
{\it something} that is consistent with the expansion model is indeed true,
when even the expansion model itself is totally in question because of
its definitively wrong matter-energy budget prediction?

Hubble's initial caution would be much more desired, and remains valid
today. He had the essential skeptical attitude of a real investigator
of Nature.

Today, we must doubt the reality of the expansion because the
expansion scenario is part of a cosmological model that has failed in
giving a consistent picture of the universe we live.

\subsection{Sandage's style }
The fragility of Sandage's scientific approach is hidden under an
extreme pedagogical style of paper writing. His copious use of
scientific references and textbook style confuses rather than
convinces the critical Reader.

It is curious --- and one is referred here to the realm of psychology
--- that Sandage does not mention the most likely and scientifically
palatable reason for Hubble's reluctance in accepting the expanding
universe explanation of his redshift-distance law: {\it the age 
problem}. With Hubble's constant of the time, the age of the universe
turns out to be about half of the geological age of the Earth. Hubble
died in 1953, precisely when Walter Baade made the first substantial
revision of Hubble's constant. History tells us then that Sandage
himself devoted a gigantic effort to put it even down, reaching
finally the now famous 50 figure. One might well speculate --- in the
realm of psychology still --- that Sandage does not mention the age
problem as the main scientific reason for Hubble's doubt because he
would be revealing his {\it own personal hell}: he fights also with an age
problem --- remember, he is a celebrated champion of modern cosmology
--- and that is the reason of his beloved 50 or lower.

\subsection{Last  }
The age problem, again and again. Where has it led modern Big Bang
cosmology to? To {\it a completely dark and unknown universe.} But, in
principle, that is not a big problem at all, as long as one is
satisfied with playing with universe toy-models. Exactly the way we
witness Sandage and collaborators doing with their investigation of
the Tolman effect.

\subsection{But not least }
A. Brynjolfsson (2006) discussed Lubin and Sandage's data in the light of 
plasma redshift theory. He claims that the Tolman test is consistent with 
plasma redshift cosmology (Brynjolfsson 2004) which predicts that the Tolman 
factor is close to (1+z)$^3$ and not to (1+z)$^4$, as required by the 
Big-Bang cosmology. It is worthwhile to reproduce the abstract of
Brynjolfsson's 2006 work mentioned above.
\begin{quote}
``Surface Brightness Test and Plasma Redshift"\\
The plasma redshift of photons in a hot sparse plasma follows from basic 
axioms of physics. It has no adjustable parameters (arXiv:astro-ph/0406437). 
Both the distance-redshift relation and the magnitude-redshift relation for 
supernovae and galaxies are well-defined functions of the average electron 
densities in intergalactic space. We have previously shown that the predictions 
of the magnitude-redshift relation in plasma- redshift cosmology match well 
the observed relations for the type Ia supernovae (SNe). No adjustable 
parameters such as the time variable ``dark energy'' and ``dark matter'' 
are needed. We have also shown that plasma redshift cosmology predicts well the 
intensity and black body spectrum of the cosmic microwave background (CMB). 
Plasma redshift explains also the spectrum below and above the 2.73 K black body 
CMB, and the X-ray background. In the following, we will show that the good 
observations and analyses of the relation between surface brightness and 
redshift for galaxies, as determined by Allan Sandage and Lori M. Lubin in 2001, 
are well predicted by the plasma redshift. All these relations are 
inconsistent with cosmic time dilation and the contemporary big-bang 
cosmology. 
\end{quote}

C.F. Gallo (2006) presented, in the 2006 April meeting of the American 
Physical Society, work in progress, in which he discusses a  general 
thermodynamic argument that would justify a {\it ``Tired Light Concept"}. In order 
to duplicate a Doppler Redshift it is required a detailed microscopic treatment 
of the photon/light interaction with the interacting medium (plasma, atoms, 
molecules, negative ions, etc), which has not been conclusively demonstrated 
theoretically or experimentally yet. 

Gallo's abstract presented at the APS meeting is reproduced below.
\begin{quote}
``Thermalization Tendency of Electromagnetic Radiation in Transit Through 
Astrophysical Mediums"\\
As Electromagnetic Radiation from a hot source transits through a cooler 
interacting medium, the following are demonstrated from thermodynamic 
arguments. \\
(1) The ``hot" radiation always loses some energy to the cooler 
interacting medium. \\
(2) Detailed behavior depends upon the microscopic nature 
of the interacting medium. \\
(3) A Redshift will occur, but not necessarily 
imitate the wavelength dependence of the Doppler Redshift. \\
(4) A Doppler-type redshift will occur only under certain conditions.\\
(5) The loss of radiative energy to the 
intergalactic medium will contribute to the Cosmic Microwave Background 
Radiation. \\
The following characteristics depend upon the detailed nature of 
the interacting medium. \\
(1) The photon energy loss per collision. \\
(2) The magnitude (cross-sections) of the thermalization process. \\
(3) The energy dependence of the cross- section for various mediums. \\
(4) Forward propagation characteristics of the Redshifted EM radiation. \\
Although the effects are small, the cumulative redshift in astrophysical 
situations can be significant. 
Earthly experiments are planned. 
\end{quote}
At this point it is interesting to recall what happened in the past, in a similar 
situation, when Einstein gave a heuristic interpretation to the {\it photoelectric 
effect}. One can make an useful counterpoint to the {\it redshift effect} 
observed by Hubble. 

Einstein's heuristic model departed from the following experimental evidences 
(e.g., Stachel 1998): 
\begin{description}
\item{(a)} the effect does not depend on the intensity of the radiative source;
\item{(b)} short wavelength blackbody radiation is described by the Wien limit;
\item{(c)} large wavelength blackbody radiation is described by the Rayleigh-Jeans 
distribution.
\end{description}
A heuristic program for the redshift effect might likewise consider at least 
the following observational evidences:
\begin{description}
\item{(a)} the effect depends on the flux of the source according to Hubble's law;
\item{(b)} the effect does not depend on the wavelength of the radiation;
\item{(c)} the effect is quantized (Tifft 2003, Arp 1998 and references therein).
\end{description}
Such a program would certainly clear the way for a theory to the tired-light 
paradigm.

Turning now to the Microwave Background Radiation (MBR), Halton Arp in one of his 
books (Arp 1998, p. 237) cites an authentic Fred Hoyle's aphorism: 
\begin{quote}
{\it
``A man who falls asleep on the top of a mountain and who awakes in a fog does not 
think he is looking at the origin of the Universe. He thinks he is in a fog."}
\end{quote}
Let us then consider a {\it local} approach to MBR. Being freed from the 
``prison" of the Hot Big Bang Cosmology one may speculate on an earthly origin 
for the MBR. Earth's magnetosphere can be seen as a {\it magnetic bottle} whose 
walls are made by solar wind particles trapped along the magnetic lines of the 
Earth field. A minute fraction of Sun's light reflected by the Earth 
surface is caught within such a bottle and is thermalized through Thomson 
scattering on the bottle walls. The first consequence is that one would expect 
that the thermalized radiation should exhibit a {\it dipole anisotropy}, 
given the nature of Earth's  magnetic field. And that is precisely what was 
observed by the COBE satellite from its 900-km altitude orbit.   

\begin{figure}[H]
\begin{center}
\includegraphics[width=10cm]{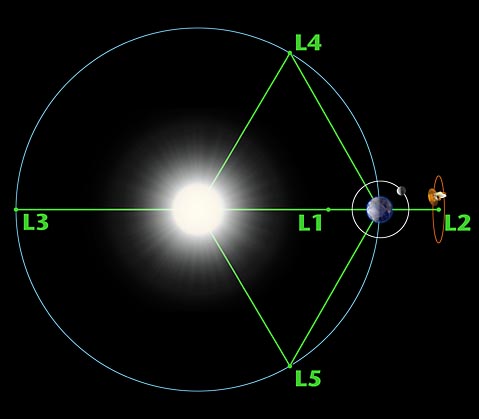}
\end{center}
\vskip 0.4cm
\caption{\small Lagrangean points of the Sun-Earth system. WMAP is shown around L2.  
Image credit: Wilkinson Microwave Anisotropy Probe electronic page. }
\label{fig:f16-1}
\end{figure}

A straight consequence --- easily testable --- 
is that the background radiation from other ``magnetic bottles" --- other 
planets --- will be different, with a different thermal spectrum, possibly 
non thermal and even nonexistent. A probe orbiting another solar system planet 
like Mars, Venus, etc, would verify the hypothesis. Although WMAP, the 
{\it Wilkinson Microwave Anisotropy Probe}, sits far away 
from Earth, at the Lagrangean L2 point of the Sun-Earth system (see WMAP 
electronic page at the URL 
\url{http://map.gsfc.nasa.gov/m\_mm/ob\_techorbit1.html}), which means about 1.5 
million km from Earth, that is not enough for it to be released from the magnetic 
influence from Earth (Figure \ref{fig:f16-1}). 

\begin{figure}[H]
\begin{center}
\includegraphics[width=10cm]{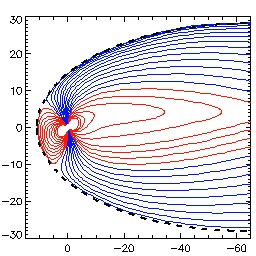}
\end{center}
\vskip 0.4cm
\caption{\small A view of Earth's magnetopause. The 
bullet-shaped magnetopause is always along the Sun-Earth direction. L2 is 
inside the magnetopause at about 230 Earth radii. 
Image credit: ``The Exploration of the Earth's Magnetosphere", an educational 
web site by David P. Stern and Mauricio Peredo. }
\label{fig:f16-2}
\end{figure}

Although its large altitude, it is located precisely 
and deep inside the bullet-shaped magnetopause, which extends to 1000 
times the Earth radius or more --- approximately 10 million km (see Figure \ref{fig:f16-2} and \url{http://www-spof.gsfc.nasa.gov/Education/wmpause.html} for more details 
about the magnetopause). 

\section{References}
\begin{description}
\item Arp, H. 1998, {\it Seeing Red: Redshifts, Cosmology and Academy,} Apeiron, 
Montreal
\item Brynjolfsson, A. 2004, arXiv:astro-ph/0406437
\item Brynjolfsson, A. 2006, APS Joint Spring Meeting of the New England Section, 
March 31-April 1, 2006, Boston, Massachusetts, abstract \#B.00008 
\item Gallo, C.F. 2006, American Physical Society, April Meeting, April 22-25, 
2006, Dallas, Texas, abstract \#J7.00007
\item Hoyle, F. 2001, in {\it Universe, The Cosmology Quest}, DVD directed 
by Randall Meyers, A Floating World Films production
\item Lubin, L.M. \& Sandage, A. 2001, AJ, 122, 1084 (LS01, \\
arXiv:astro-ph/0106566) 
\item Pecker, J.-C. 2001, in {\it Universe, The Cosmology Quest}, DVD directed 
by Randall Meyers, A Floating World Films production
\item Soares, D.S.L. 2002, {\it Do we live in an anthropic universe?},\\  
arXiv:physics/0209094 
\item Stachel, J. 1998, (org.) {\it Einstein's Miraculous Years: Five Papers 
that Changed the Face of Physics,} Princeton University Press, Princeton
\item Tifft, W.G. 2003, Ap\&SS, 285, 429
\item Tolman, R.C. 1930, Proc. Natl. Acad. Sci., 16, 511 
\end{description}

\end{document}